\documentstyle[12pt]{article}
\newcommand{\lpar}{\stackrel{\leftarrow}{\partial}}
\newcommand{\rpar}{\stackrel{\rightarrow}{\partial}}
\begin{document}
\renewcommand{\thefootnote}{\fnsymbol{footnote}}
\begin{center}
{\large\bf Linear Odd Poisson Bracket  \\
on Grassmann Variables \\ }
\vspace{1cm} Vyacheslav A. Soroka
\footnote{E-mail:  vsoroka@kipt.kharkov.ua}
\vspace{1cm}\\
{\it Institute of Theoretical Physics}\\
{\it National Science Center}\\
{\it "Kharkov Institute of Physics and Technology"}\\
{\it 310108, Kharkov, Ukraine}\\
\vspace{1.5cm}
\end{center}
\begin{abstract}
A linear odd Poisson bracket (antibracket) realized
solely in terms of Grassmann variables is suggested. It is revealed
that the bracket, which corresponds to a semi-simple Lie group, has at
once three Grassmann-odd nilpotent $\Delta$-like differential operators of
the first, the second and the third orders with respect to Grassmann
derivatives, in contrast with the canonical odd Poisson bracket having
the only Grassmann-odd nilpotent differential $\Delta$-operator of the
second order. It is shown that these $\Delta$-like operators together
with a Grassmann-odd nilpotent Casimir function of this bracket form a
finite-dimensional Lie superalgebra.

\bigskip
\noindent
{\it PACS:} 02.20.Sv; 11.15.-q; 11.30.Pb

\medskip
\noindent
{\it Keywords:} Poisson bracket; Lie group; Lie superalgebra
\end{abstract}
\renewcommand{\thefootnote}{\arabic{footnote}}
\setcounter{footnote}0

\newpage
{\bf 1.\/} Recently a linear degenerate odd Poisson bracket built only of
Grassmann variables has been introduced \cite{S}. It was constructed for
this bracket, in contrast with the non-degenerate odd bracket having the
only Grassmann-odd nilpotent differential $\Delta$-operator of the second
order, at once three Grassmann-odd nilpotent $\Delta$-like
differential operators of the first, the second and the third orders with
respect to Grassmann derivatives. It was also shown that these
$\Delta$-like operators together with a Grassmann-odd nilpotent Casimir
function of this degenerate odd bracket form a finite-dimensional Lie
superalgebra. In the present paper we extend the above-mentioned results
to the case of an arbitrary linear odd Poisson bracket, which is also
realized solely in terms of the Grassmann variables and corresponds to a
semi-simple Lie group.

{\bf 2.\/} There is a well-known linear even Poisson bracket given
in terms of the commuting (Grassmann-even) variables $X_\alpha$
$(g(X_\alpha) = 0)$
$$
\{X_\alpha, X_\beta \}_0 = {c_{\alpha\beta}}^\gamma X_\gamma\ ,\qquad
(\alpha,\beta,\gamma = 1,...,N),
\eqno {(1)}
$$
where ${c_{\alpha\beta}}^\gamma$ are Grassmann-even
$(g({c_{\alpha\beta}}^\gamma) = 0)$ constants which, because of the main
properties of the even Poisson bracket:
$$
\{ A, B + C \}_0 = \{ A, B \}_0 + \{ A, C \}_0\ ,
\eqno {(2)}
$$
$$
g(\{ A, B \}_0) = g(A) + g(B) \pmod 2\ ,
\eqno {(3)}
$$
$$
\{ A , B \}_0 = -(-1) ^{g(A) g(B)} \{ B , A \}_0\ ,
\eqno {(4)}
$$
$$
\sum_{(ABC)}(-1) ^{g(A) g(C)} \{ A , \{ B , C \}_0 \}_0 = 0\ ,
\eqno {(5)}
$$
$$
\{ A , B C \}_0 = \{ A , B \}_0\ C +
(- 1) ^{g(A) g(B)}\ B \{ A , C \}_0\ ,
\eqno {(6)}
$$
are antisymmetric in the two lower indices
$$
{c_{\alpha\beta}}^\gamma = - {c_{\beta\alpha}}^\gamma
\eqno {(7)}
$$
and obey the conditions
$$
{c_{\alpha\lambda}}^\delta {c_{\beta\gamma}}^\lambda +
{c_{\beta\lambda}}^\delta {c_{\gamma\alpha}}^\lambda +
{c_{\gamma\lambda}}^\delta {c_{\alpha\beta}}^\lambda = 0\ .
\eqno {(8)}
$$
A sum with the symbol $(ABC)$ in (5) means a summation over cyclic
permutations of the quantities $A, B, C$. In relations (2)-(6) $A, B, C$
are functions of the variables $X_\alpha$ and $g(A)$ is a Grassmann parity
of the quantity $A$. The linear even bracket (1) plays a very important
role in the theory of Lie groups, Lie algebras, their representations and
applications (see, for example, \cite{b,km}).

As in the Lie algebra case, we can define a symmetric Cartan-Killing
tensor
$$
g_{\alpha\beta} = g_{\beta\alpha} =
{c_{\alpha\gamma}}^\lambda {c_{\beta\lambda}}^\gamma
\eqno {(9)}
$$
and verify with the use of relations (8) an anti-symmetry property of a
tensor
$$
c_{\alpha\beta\gamma} = {c_{\alpha\beta}}^\delta g_{\delta\gamma} =
- c_{\alpha\gamma\beta}\ .
\eqno {(10)}
$$
By assuming that the Cartan-Killing metric tensor is non-degenerate
$det(g_{\alpha\beta}) \neq 0$ (this case corresponds to the semi-simple
Lie group), we can define an inverse tensor $g^{\alpha\beta}$ $$
g^{\alpha\beta}g_{\beta\gamma} = \delta^\alpha_\gamma\ ,
\eqno {(11)}
$$
with the help of which we are able to build a quantity
$$
C = X_\alpha X_\beta g^{\alpha\beta}\ ,
$$
that, in consequence of relation (10), is a Casimir function for the
bracket (1)
$$
\{ C, ...\}_0 = 0\ .
$$

{\bf 3.\/} Now let us replace in expression (1) the commuting variables
$X_\alpha$ by Grassmann variables $\theta_\alpha$ $(g(\theta_\alpha) = 1)$.
Then we obtain a binary composition
$$
\{\theta_\alpha, \theta_\beta \}_1 =
{c_{\alpha\beta}}^\gamma \theta_\gamma\ ,
\eqno {(12)}
$$
which, due to relations (7) and (8), meets all the properties of the odd
Poisson brackets:
$$
\{ A, B + C \}_1 = \{ A, B \}_1 + \{ A, C \}_1\ ,
\eqno {(13)}
$$
$$
g(\{ A, B \}_1) = g(A) + g(B) + 1 \pmod 2\ ,
\eqno {(14)}
$$
$$
\{ A , B \}_1 = -(-1) ^{(g(A) + 1)(g(B) + 1)} \{ B , A \}_1\ ,
\eqno {(15)}
$$
$$
\sum_{(ABC)}(-1) ^{(g(A) + 1)(g(C) + 1)} \{ A , \{ B , C \}_1 \}_1 = 0\ ,
\eqno {(16)}
$$
$$
\{ A , B C \}_1 = \{ A , B \}_1\ C +
(- 1) ^{(g(A) + 1)g(B) }\ B \{ A , C \}_1\ .
\eqno {(17)}
$$
It is surprising enough that the odd bracket can be defined solely
in terms of the Grassmann variables as well as an even Martin bracket
\cite{mar}.

By the way, let us note that we can also construct only on the
Grassmann variables a non-linear odd Poisson bracket of the form
$$
\{\theta_\alpha, \theta_\beta \}_1 =
{c_{\alpha\beta}}^{\gamma\delta\lambda}
\theta_\gamma \theta_\delta \theta_\lambda\ .
$$
In order to satisfy the property (15), the constants
${c_{\alpha\beta}}^{\gamma\delta\lambda}$ have to be antisymmetric in the
below indices
$$
{c_{\alpha\beta}}^{\gamma\delta\lambda} =
- {c_{\beta\alpha}}^{\gamma\delta\lambda}
$$
and, for the validity of the Jacobi identities (16), they must obey the
following conditions
$$
\sum_{(\alpha\beta\gamma)} {c_{\alpha\beta}}^{\lambda[\delta_1\delta_2}
{c_{\lambda\gamma}}^{\delta_3\delta_4\delta_5]} = 0\ ,
$$
where square brackets $[...]$ mean the antisymmetrization of the indices in
them.

Returning to the linear odd bracket (12) notice, that on functions $A, B$
of Grassmann variables $\theta_\alpha$ it has the form
$$
\{ A , B \}_1 = A \lpar_{\theta_\alpha}
{c_{\alpha\beta}}^\gamma \theta_\gamma \rpar_{\theta_\beta} B\ ,
$$
where $\lpar$ and $\rpar$ are the right and left derivatives and
$\partial_{x^A} \equiv {\partial \over {\partial x^A}}$.
The bracket (12) can be either degenerate or non-degenerate in the
dependence on whether the matrix ${c_{\alpha\beta}}^\gamma \theta_\gamma$
in the indices $\alpha, \beta$ is degenerate or not. The indices
$\alpha, \beta$ can be raised and lowered by means of the
non-degenerate metric tensors (9), (11)
$$
\theta^\alpha = g^{\alpha\beta} \theta_\beta\ ,\qquad
\partial_{\theta^\alpha} =
g_{\alpha\beta} \partial_{\theta_\beta}\ .
$$
Hereafter only the non-degenerate metric tensors (11) will be considered.

{\bf 4.\/} By contracting the indices in a product of the three Grassmann
variables with the upper indices $\theta^\alpha \theta^\beta \theta^\gamma$
and of the three successive Grassmann derivatives
$\partial_{\theta_\alpha} \partial_{\theta_\beta} \partial_{\theta_\gamma}$,
respectively, with the lower indices $\alpha, \beta, \gamma$ in (8),
we obtain the relations
$$
\theta^\alpha \theta^\beta \theta^\gamma
{c_{\alpha\beta}}^\lambda {c_{\lambda\gamma}}^\delta = 0\ ,
\eqno {(18)}
$$
$$
{c_{\alpha\beta}}^\lambda {c_{\lambda\gamma}}^\delta
\partial_{\theta_\alpha} \partial_{\theta_\beta}
\partial_{\theta_\gamma} = 0\ ,
\eqno {(19)}
$$
which will be used later on many times. In particular, taking into account
relation (18), we can verify that the linear odd bracket (12) has the
following Grassmann-odd nilpotent Casimir function
$$
\Delta_{+3} = {1\over\sqrt{3!}} \theta^\alpha \theta^\beta \theta^\gamma
c_{\alpha\beta\gamma}\ ,\qquad
\{\Delta_{+3}, ...\}_1 = 0\ ,\qquad(\Delta_{+3})^2 = 0\ .
\eqno {(20)}
$$
Such a notation for this Casimir function will be clear below.

It is a well-known fact that, in contrast with the even Poisson bracket,
the non-degenerate  odd Poisson bracket has one Grassmann-odd nilpotent
differential $\Delta$-operator of the second order, in terms of which the
main equation has been formulated in the Batalin-Vilkovisky scheme
\cite{bv,bv1,blt,bt,sc,kn} for the quantization of gauge theories in
the Lagrangian approach. In a formulation of Hamiltonian dynamics
by means of the odd Poisson bracket with the help of a Grassmann-odd
Hamiltonian $\bar H$ $(g(\bar H) = 1)$
\cite{l,vpst,s,k,kn1,vst,vs,s1,s2,n,vty,s3} this
$\Delta$-operator plays also a very important role being used to
distinguish the non-dissipative dynamical systems, for which
$\Delta \bar H = 0$, from the dissipative ones \cite{S}, for which the
Grassmann-odd Hamiltonian satisfies the condition $\Delta \bar H \neq 0$.

Now let us try to build the $\Delta$-operator for the linear
odd bracket (12). It is remarkable that, in contrast with the
canonical odd Poisson bracket having the only $\Delta$-operator of the
second order, we are able to construct at once three $\Delta$-like
Grassmann-odd nilpotent operators which are differential operators of
the first, the second and the third orders respectively
$$
\Delta_{+1} = {1\over\sqrt{2}} \theta^\alpha \theta^\beta
c_{\alpha\beta\gamma} \partial_{\theta_\gamma}
\ ,\qquad (\Delta_{+1})^2 = 0\ ;
\eqno {(21)}
$$
$$
\Delta_{-1} = {1\over\sqrt{2}} \theta_\gamma
{c_{\alpha\beta}}^\gamma
\partial_{\theta_\alpha} \partial_{\theta_\beta}
\ ,\qquad (\Delta_{-1})^2 = 0\ ;
\eqno {(22)}
$$
$$
\Delta_{-3} = {1\over\sqrt{3!}} c_{\alpha\beta\gamma}
\partial_{\theta_\alpha} \partial_{\theta_\beta} \partial_{\theta_\gamma}
\ ,\qquad (\Delta_{-3})^2 = 0\ .
\eqno {(23)}
$$
The nilpotency of the operators $\Delta_{+1}$ and $\Delta_{-1}$ is a
consequence of relations (18) and (19).

It is also interesting to reveal that these $\Delta$-like operators
together with the Casimir function $\Delta_{+3}$ (9) are closed into the
finite-dimensional Lie superalgebra, in which the anticommuting relations
between the quantities $\Delta_\lambda$ $(\lambda = -3, -1, +1, +3 )$
(20)-(23) with the nonzero right-hand side  are\footnote{In order to
avoid a misunderstanding, let us note that below $[A, B] = AB -BA$ and $\{
A, B\} = AB + BA$.}
$$
\{ \Delta_{-1}, \Delta_{+1} \} = Z\ ,
\eqno {(24)}
$$
$$
\{ \Delta_{-3}, \Delta_{+3} \} = N - 3Z\ ,
\eqno {(25)}
$$
where
$$
N = - c^{\alpha\beta\gamma} c_{\alpha\beta\gamma}
$$
is a number of values for the indices $\alpha, \beta, \gamma$
$(\alpha, \beta, \gamma = 1,...,N)$ and
$$
Z = D - K
\eqno {(26)}
$$
is a center of this superalgebra
$$
[ Z, \Delta_\lambda ] = 0\ ,\qquad (\lambda = -3, -1, +1, +3)\ .
\eqno {(27)}
$$
In (26)
$$
D = \theta_\alpha \partial_{\theta_\alpha}
\eqno {(28)}
$$
is a "dilatation" operator for the Grassmann variables
$\theta_\alpha$ which distinguishes the $\Delta_\lambda$-operators
with respect to their uniformity degrees in $\theta$
$$
[ D, \Delta_\lambda ] = \lambda
\Delta_\lambda\ ,\qquad (\lambda = -3, -1, +1, +3)\ ,
\eqno {(29)}
$$
and the quantity $K$ has the form
$$
K = {1\over 2} \theta^\alpha \theta^\beta
{c_{\alpha\beta}}^\lambda c_{\lambda\gamma\delta}
\partial_{\theta_\gamma} \partial_{\theta_\delta}\ .
\eqno {(30)}
$$
The operator $Z$ is also a center of the Lie superalgebra which
contains both the operators $\Delta_\lambda$ (20)-(23), $Z$ (26) and the
operator $D$ (28)
$$
[Z, D] = 0\ .
\eqno {(31)}
$$

We can add to this superalgebra the generators
$$
S_\alpha = \theta_\gamma {c_{\alpha\beta}}^\gamma
\partial_{\theta_\beta}
\eqno {(32)}
$$
with the following commutation relations:
$$
[ S_\alpha, S_\beta ] = {c_{\alpha\beta}}^\gamma S_\gamma\ ,\qquad
[ S_\alpha, \Delta_\lambda ] = 0\ ,
\eqno {(33a,b)}
$$
$$
[ S_\alpha, Z ] = 0\ ,\qquad [ S_\alpha, D ] = 0\ .
\eqno {(33c,d)}
$$
In order to prove the permutation relations for the Lie superalgebra
(20)-(33), we have to use relations (18) and (19).
Note that the center $Z$ (26) coincides with the expression for a
quadratic Casimir operator of the Lie algebra (33a) for the generators
$S_\alpha$ given in the representation (32)
$$
S_\alpha S_\beta g^{\alpha\beta} = Z\ .
\eqno {(34)}
$$

{\bf 5.\/} Thus, we see that both the even and odd linear Poisson brackets
are internally inherent in the Lie group with the structure constants
subjected to conditions (7) and (8).  However, only for the linear odd
Poisson bracket realized in terms of the Grassmann variables and only in
the case when this bracket corresponds to the semi-simple Lie group,
there exists the Lie superalgebra (20)-(33) for the $\Delta$-like
operators of this bracket.

Note that in the case of the degenerate Cartan-Killing
metric tensor (9), relation (10) remains valid and we can construct only
two $\Delta$-like Grassmann-odd nilpotent operators: $\Delta_{-1}$ (22)
and $\Delta_{-3}$ (23), which satisfy the trivial anticommuting relation
$$
\{ \Delta_{-1}, \Delta_{-3} \} = 0\ .
$$

The Lie superalgebra (20)-(33), naturally connected with the linear odd
Poisson bracket (12), may be useful for the subsequent development of the
Batalin-Vilkovisky formalism for the quantization of gauge theories.
Indeed, very similar to (12) odd Poisson brackets on the Grassmann algebra
are used in a generalization \cite{g}\footnote{This paper appeared after
the present work had been finished.} of the triplectic formalism
\cite{bms} which is a covariant version of the $Sp(2)$-symmetric
quantization \cite{blt} of general gauge theories. We should therefore
expect that the Lie superalgebra (20)-(33), closely related with the
linear odd bracket (12), will also find the application for the further
development of the above-mentioned generalization of the triplectic
formalism. Let us note that the superalgebra (20)-(33) can also be used
in the theory of representations of the semi-simple Lie groups.

\bigskip
The author is sincerely thankful to V.D. Gershun for useful discussions.

\medskip
This work was supported, in part, by the Ukrainian State Foundation
of Fundamental Researches, Grant No 2.5.1/54, by Grant INTAS No 93-127
(Extension) and by Grant INTAS No 93-633 (Extension).

\end{document}